\documentclass{aa}  

\usepackage{graphicx}
\usepackage{natbib}
\usepackage{ulem}
\usepackage{longtable}
\usepackage{subfig}
\usepackage{float}
\usepackage[usenames,dvipsnames]{xcolor}
\usepackage{multirow}
\usepackage{footnote}
\usepackage[bottom]{footmisc}
\usepackage{ulem}

\usepackage{txfonts}

\newcommand{\teff}{T_\mathrm{eff}}
\newcommand{\logg}{\mathrm{log}\,g}

\begin{document} 

   \title{Spectral energy distribution of M-subdwarfs: A study of their atmospheric properties 
    \thanks{Based on observations made with the ESO Very Large Telescope at the Paranal Observatory under programme 092.D-0600(A)}}

\titlerunning{Spectral energy distribution of M-subdwarfs}
 \authorrunning{Rajpurohit et al}

\author{Rajpurohit, A. S.\inst{1,2}, Reyl\'e, C.\inst{3}, Allard, F.\inst{4}, Homeier, D.\inst{5,}\inst{4}, Bayo, A.\inst{6}, Mousis, O.\inst{1}, Rajpurohit, S. \inst{7}, Fern\'andez-Trincado, J. G.\inst{3}}

\institute{Astronomy \& Astrophysics Division, Physical Research Laboratory, Ahmedabad 380009, India
\email{arvindr@prl.res.in}
\and
Aix Marseille Universit\'{e}, CNRS, LAM (Laboratoire d'Astrophysique 
de Marseille) UMR 7326, 13388, Marseille, France 
\and
Institut UTINAM CNRS 6213, Observatoire des Sciences de l'Univers THETA Franche-
Comt\'{e} Bourgogne, Univ. Bourgogne Franche-Comt\'{e}, 
Observatoire de Besan\c{c}on, BP 1615, 25010 Besan\c{c}on Cedex, France
\and
Univ Lyon, Ens de Lyon, Univ Lyon1, CNRS, 
Centre de Recherche Astrophysique de Lyon UMR5574, 
F-69007, Lyon, France
\and
Zentrum f{\"u}r Astronomie der Universit{\"a}t Heidelberg, Landessternwarte,
K\"{O}nigstuhl 12, 69117 Heidelberg, Germany
\and
Instituto de F\'isica y Astronom\'ia, Facultad de Ciencias, Universidad de Valpara\'iso, Av. Gran Breta\~na 1111, 5030 Casilla, Valpara\'iso, Chile
\and
Clausthal University of Technology, Institute for Theoretical Physics, Leibnizstr.10, 38678 Clausthal-Zellerfeld, Germany
}

   \date{Received May 1, 2016; accepted }

 
 \abstract
    {M-type subdwarfs are metal-poor low-mass stars and are probes for the old populations in our Galaxy.  Accurate knowledge of their atmospheric parameters and  especially their composition is essential for understanding the chemical history of our Galaxy.
    }
   {The purpose of this work is to perform a detailed study of M-subdwarf spectra covering the full wavelength range from the optical to the near-infrared. It allows us to perform a more detailed analysis of the atmospheric composition in order to determine the stellar parameters, and to constrain the atmospheric models. The study will allow us to further understand physical and chemical processes such as increasing condensation of gas into dust, to point out the missing continuum opacities, and to see how the main band features are reproduced by the models. The spectral resolution and the large wavelength coverage used is a unique combination  that can constrain the processes that occur in a cool atmosphere.}  
   {We obtained medium-resolution spectra (R = 5000-7000) over the wavelength range 0.3-2.5 $\mu$m of ten M-type subdwarfs with X-SHOOTER at VLT.  These data constitute a unique atlas of M-subdwarfs from optical to near-infrared. We performed a spectral synthesis analysis using a full grid of synthetic spectra computed from BT-Settl models and obtained consistent stellar parameters such as effective temperature, surface gravity, and metallicity.}
   {We show that state-of the-art atmospheric models correctly represent the overall shape of their spectral energy distribution, as well as atomic and molecular line profiles both in the optical and near-infrared. We find that the actual fitted gravities of almost all our sample are consistent with old objects, except for LHS 73 where it is found to be surprisingly low.
  
   }
   {}

   \keywords{Stars: low-mass -- subdwarfs -- atmospheres}

   \maketitle
%

\section{Introduction}
M-type stars are the most common stars in our Galaxy \citep[70$\%$ of the Galactic stellar population;][]{Bochanski2010}. They are an important probe for our Galaxy as they carry fundamental information regarding the composition history, a record of galactic structure and formation, and of its dynamics. Although they are not expected to be as numerous  as compared to M-dwarfs \citep[0.25$\%$ of the Galactic stellar population;][] {Reid2005b}, the metal-poor M-type subdwarfs can serve as probes for the old populations in our Galaxy (old disc, thick disc and halo). The locus of M-subdwarfs (sdMs) in the Hertzsprung-Russel (HR) diagram deviates from most of the field stars owing to the metallicity difference that translates into different opacities from those of regular M-dwarfs. Since some subdwarfs lie close to the hydrogen burning limit, they can be used to probe the lower end of the stellar mass function, extending it towards the brown dwarf regime. Because of their intrinsic faintness and the difficulty in getting a  homogeneous sample of unique age and metallicity, very little is known about them. As we go from earlier to later M-subtypes, more molecules form in their atmospheres, making the spectral continuum very hard or impossible to identify both in the optical and in the near-infrared (NIR). Furthermore they provide the right condition for studying molecules and dust  formation in the low-metallicity environment as well as radiative transfer in cool, metal-poor atmospheres.

With decreasing temperature, sdMs spectra show an increase in abundances of diatomic and triatomic molecules in the optical and in the near-infrared (e.g. SiH, CaH, CaOH, TiO, VO, CrH, FeH, OH, H$_2$O, CO). The molecules TiO  and VO dominate the opacity sources in the optical and  H$_2$O and CO at lower spectral resolution in the infrared, having more complex and extensive band structures which leave no window for the true continuum and create a pseudo-continuum that at low spectral resolution only shows the strongest, often resonant atomic lines \citep{Allard1990,Allard1995}. However, owing to their low metallicity,  and hence low Ti and O abundances \citep{Savcheva2014}, the TiO bands are not as strong, and the pseudo-continuum is brighter as a result. This increases the contrast with the other opacities, such as hydride bands and atomic lines that feel the higher pressures of the deeper layers from where they emerge \citep{Allard1990,Allard1995}. 

Different classification schemes have been proposed to assign metallicity and spectral types of sdMs. \cite{Ryan1991a} and \cite{Ryan1991b} used metallic  lines, such as the CaII K line, to determine the metallicity of subdwarfs. \cite{Baraffe1997} published the first evolution tracks and isochrones based of the NextGen model atmosphere \citep{Allard1997,Allard2000,Hauschildt1999} spanning the complete range of composition of (s)dMs stars. \cite{Gizis1997} proposed a first classification of sdMs and extreme subdwarfs (esdMs) based on the NextGen isochrone luminosity and TiO and CaH band strengths in low-resolution optical spectra. \cite{Lepine2007} has revised the adopted classification and proposed a new classification for the most metal poor, the ultra subdwarfs (usdMs). \cite{Jao2008} has compared his model grids with the optical spectra to characterise the spectral energy distribution of subdwarfs using three parameters-- temperature, gravity and metallicity-- and thus gave an alternative classification scheme of subdwarfs. 

The proper classification of these objects requires the grid of synthetic spectra to be compared with the observations, which then help to quantify their basic properties: elemental abundances, effective temperature, and surface gravity. At present, these physical properties are not yet particularly well determined for sdMs. Traditional techniques to estimate stellar effective temperature based on black-body approximations and broadband photometry are at best dangerous for cool M-dwarfs whose true continua are masked by extensive molecular absorption. Furthermore, the complexity of the stellar atmosphere increases significantly with decreasing  effective temperature as dust cloud formation occurs \citep{Tsuji1996a,Tsuji1996b,Allard1998}. This is revealed by the weakening of condensible bearing opacities such as TiO, VO, CaH,and CaOH bands in the optical wavelengths by dust Rayleigh scattering, and a reddening of the infrared spectral energy distribution with weakening water bands due to dust backwarming or the greenhouse effect \citep{Allard2001}.

Over the last decades tremendous development in the model atmospheres of cool low-mass stars, in particular M-dwarfs, has been achieved \citep{Brott2005,Helling2008a,Allard2012,Allard2013} that has boosted the number of studies deriving accurate  physical parameters of these stars both in the optical and in the near-infrared  \citep{Bayo2011,Bayo2012,Rajpurohit2012a,Rajpurohit2013,Neves2014}. \cite{Bayo2014} have shown the differences between estimating the parameters in the optical and in the near-infrared, with spectra and photometry. Thanks to the large improvement of atomic and molecular line opacities which dominate the optical and infrared spectral range of these objects and to the revision of the solar abundances by \cite{Asplund2009} and \cite{Caffau2011}, synthetic spectra such as the new BT-Settl \citep{Allard2013} has achieved major improvements in modelling these complex systems. These models now even include dust cloud formation which becomes important for cool M-dwarfs and subdwarfs and yield promising results in explaining the stellar-substellar transition \citep{Allard2013,Baraffe2015}. 

Metallicity effects on the physics of cool atmospheres of very low-mass stars have been studied theoretically \citep{Allard1990,Allard1997} and then compared with observations \citep{Leggett1996,Leggett1998,Leggett2000,Leggett2001,Burgasser2002}. Such comparisons have revealed possible inaccuracies and/or incompleteness of the opacities used in the model at the time. The spectra of M-subdwarfs show the strengthening of hydride bands (OH, FeH) and pressure induced absorptions by H$_2$ around 2 $\mu$m relative to double metal bands (TiO,VO), and the broadening of atomic lines due to the effect of gravity confirming the predictions of \cite{Allard1990} and \cite{Allard1995}. Therefore we see these molecular bands in more detail in M-subdwarfs than in M-dwarfs and under more extreme gas pressure conditions. This can help reveal the remaining inaccuracies and/or incompleteness of the opacities used in the model.

In this paper, we present a homogeneous sample of M-subdwarf X-SHOOTER spectra covering the full temperature sequence from the optical to the near-infrared, and perform a spectral synthesis analysis using a full grid of BT-Settl synthetic spectra\footnote{https://phoenix.ens-lyon.fr/Grids/BT-Settl/CIFIST2011bc}. This is necessary to determine how well the models reproduce the overall observed spectral properties. We therefore determine the atmospheric parameters from the overall spectral energy distribution (SED), i.e. the simultaneous optical and near-infrared measurements, so that variability does not include further uncertainty in our estimations. We assemble ten optical to near-infrared (0.3-2.5 \,$\mu$m) spectra of sdMs, as described in \S~\ref{obs}. Section 3 present the most recent atmosphere models that we used for comparison using a method described in \S~\ref{comp}. Section 5 gives the results of the comparison and the conclusion is given in \S~\ref{ccl}.

\section{Sample selection and observations}
\label{obs}

We obtained medium-resolution spectra (R=5000 - 7000) of our sdMs sample using the X-SHOOTER spectrograph mounted on the Very Large Telescope (VLT) at the European Southern Observatory (ESO) in Paranal, Chile. In total ten targets were observed from October 2014 to March 2015. The basic properties of these objects are given in Table~1. Our sample contains late sdK7 to sdM7 and esdM2 to usdM4.5 lying well into the temperature and metallicity range of subdwarfs given by \cite{Gizis1997}, \cite{Lepine2007}, and \cite{Leggett2000}. We drew these targets from a larger sample of sdMs for which we previously derived stellar parameters from high-resolution optical spectral \citep{Rajpurohit2014}. We complement these observations with X-SHOOTER spectra to compare the fitting results for different types of observations (spectral range, resolution) and to see how the results of the analysis vary, not only with the models, but also with the quality and completeness of the observations used.

X-SHOOTER is a single-target echelle spectrograph, which maximises the sensitivity over a broad wavelength range by splitting the spectra into three different arms: UVB, covering the wavelength  range 300-559.5 nm; VIS, covering the wavelength range 559.5-1024 nm; and NIR, covering the wavelength range 1024-2480 nm. Depending on wavelength and slit width, X-SHOOTER yields a resolving power of R=4000--14000. A unique capability of X-SHOOTER is that it simultaneously collects spectra from the near-ultraviolet to the near-infrared through its three arms.

The spectra were taken with a slit width of 1” for the UVB arm and 0.9” for both the VIS and near-infrared arm yielding a homogeneous resolving power of R = 5000-7000. The exposure times were ranged from 5 to 120 minutes depending on the magnitude of the target. The signal-to-noise ratio varied over the wavelength region according to each object's spectral energy distribution and the detector's efficiency. Data were reduced using the latest software Reflex \citep{Freudling2013} for the X-SHOOTER data, which runs with standard ESO pipeline modules. The resulting 1D spectrum from each science target was then divided by the spectra of telluric stars observed just before or after, reduced and extracted using the same pipelines.

\begin{table*}[!thbp]
	\centering
	\caption{Near-infrared photometry along with coordinates for our sample are taken from 2MASS at epochs between 1997 and 2001. References are given for spectral types. The X-SHOOTER exposure times are also given.}
	\begin{tabular}{lllllllll}
		\hline\\
		Name & $\alpha$ &$\delta$ &SpT  &J &H &K$_s$ & Exp. time&References\\
		&&&&&&&(hours)&\\
		\hline\\
		LHS 73  &	 23 43 16.6& $-$24 11 14      & sdK7		&10.11&     9.59  &9.37 &0.08&a\\
		LHS 401& 15 39 39.1& $-$55 09 10      &sdM0.5	&10.15&    9.60   &9.41  &0.10&b\\
		LHS 158& 02 42 02.8&$-$44 30 59       &sdM1		&10.43&    9.94   &9.73  &0.13&b  \\
		LHS 406& 15 43 18.4&$-$20 15 31       &sdM2		&9.78  &     9.23  &9.02  &0.14&b\\
		LHS 161& 02 52 45.6&+01 55 50          &esdM2	&11.71&    11.20  &11.0  &0.33&c\\
		LHS 541& 23 17 06.0&$-$13 50 53       &sdM3		&13.03&    12.56 &12.41&0.55&d \\
		LHS 272& 09 43 46.3&$-$17 47 07       &sdM3		& 9.62 &     9.12  &8.87  &0.14 &c\\
		LHS 375& 14 31 38.3&$-$25 25 33       &esdM4	&12.15&    11.67 &11.51 &0.30&c\\
		LHS 1032&00 11 00.8&+04 20 25         &usdM4.5	&14.34&    13.81 &13.76 &1.80&e\\
		LHS 377& 14 39 00.3&+18 39 39          &sdM7		&13.19&    12.73 &12.48 &1.00&c\\
		\hline\\
	\end{tabular}
	
	(a)  \cite{Reyle2006}
	--
	(b)  \cite{Jao2008}
	--
	(c)  \cite{Gizis1997}
	--
	(d)  \cite{Dawson2000}
	--
	(e)  \cite{Lepine2007}
	
\end{table*}

\begin{figure*}[!thbp]
\vspace*{-25mm}
	\centering
\hspace*{-2cm}	
	\includegraphics[width=16.5cm,height=21cm,angle=-90]{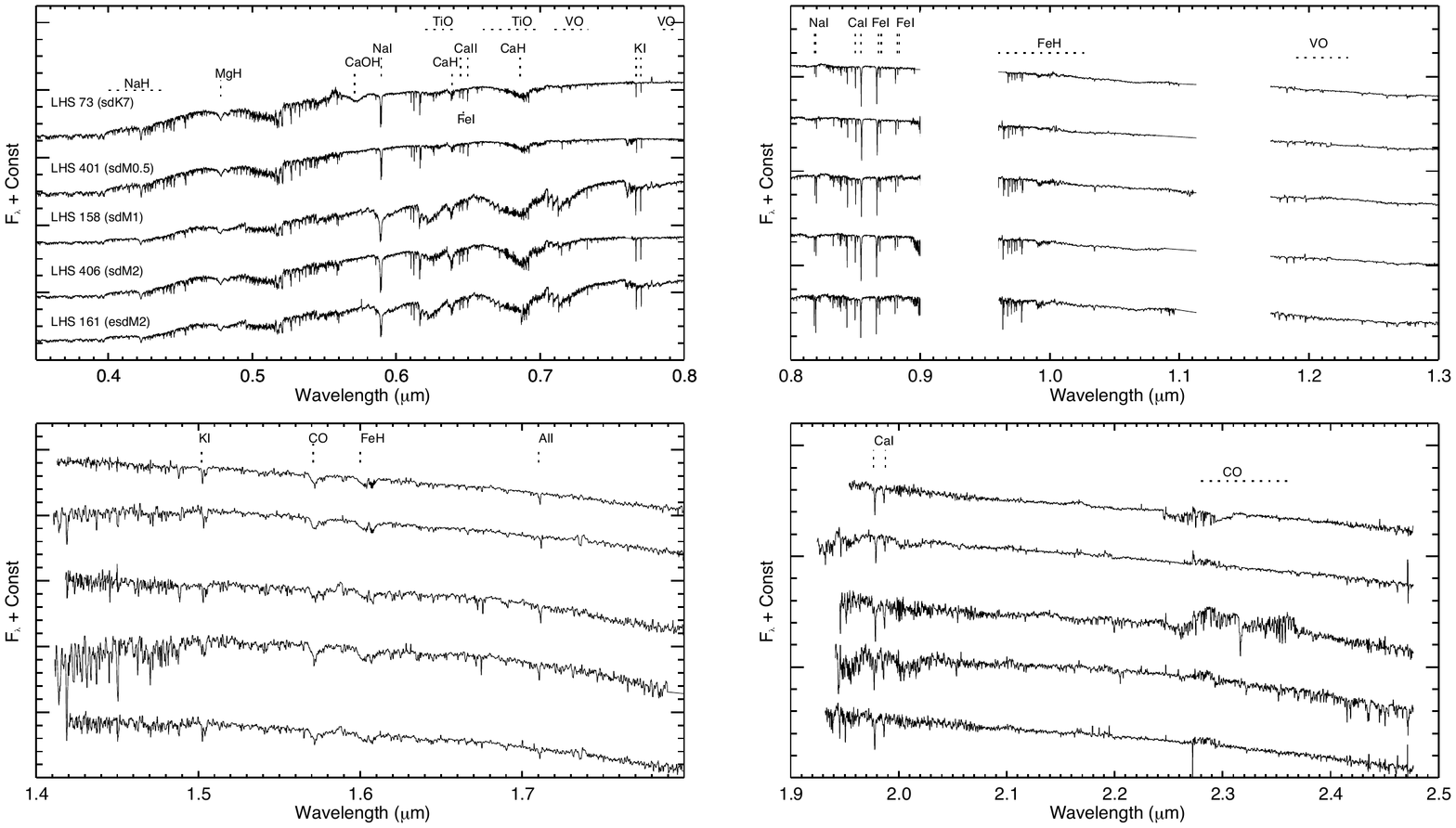}
\hspace*{-1cm}		
\vspace*{-27mm}
\caption{X-SHOOTER spectra of sdK7 to sdM2. The main spectral features are highlighted. They are mainly from \cite{Jones1994,Kirkpatrick1993,Geballe1996} and \cite{Allard1997}. The regions where telluric absorption are too strong to be properly corrected are removed from our Chi-square analysis.}
\label{Fig1}
\end{figure*}

\begin{figure*}[!thbp]
\vspace*{-25mm}
	\centering
\hspace*{-2cm}	
	\includegraphics[width=16.5cm,height=21cm,angle=-90]{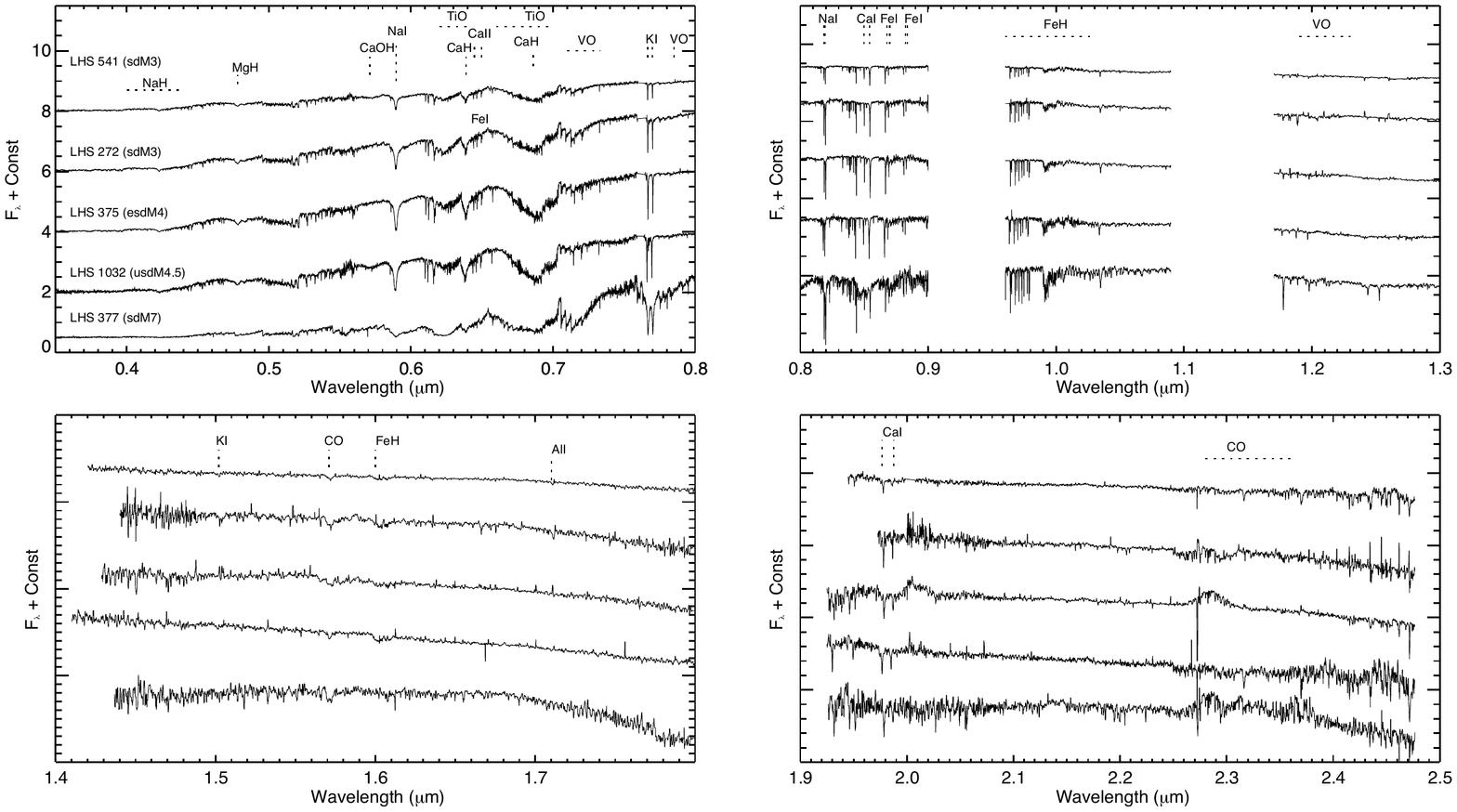}
\hspace*{-1cm}		
\vspace*{-27mm}
\caption{X-SHOOTER spectra of spectral type ranging from sdM3 to sdM7.}
\label{Fig2}
\end{figure*}

The same reduction and extraction procedures were used for the visible and NIR arms of X-SHOOTER. Given the large wavelength range covered by X-SHOOTER, seeing variation in the spectral direction can lead to wavelength-dependent slit losses. We checked the flux homogeneity of this large wavelength coverage spectral data and applied small scaling factors to each of the broadband wavelength ranges so that the spectra overlap in the common spectral regions. The scaling factor we applied ranges from less than 1$\%$ to 4$\%${,} depending on the target.

The spectral sequence of the observed subdwarfs is shown in Figure~\ref{Fig1} and Figure~\ref{Fig2}. The UVB and VIS arm spectra are dominated by molecular absorption bands from hydrides such as CaH, CaOH, NaH, and MgH. They are the most significant features since TiO is the dominant pseudo-continuum opacity in the VIS range or redder than 0.4 $\mu$m. Owing to their low metallicity, sdMs are TiO depleted. The CaH hydride is a good indicator of a subdwarf nature as it becomes very strong relative to the TiO bands. Atomic features such as Ca~II, K~I, Na~I, and FeH are clearly visible and are prominent in most of the spectra. The NIR arm spectra are dominated by various temperature sensitive species, such as the CO band, at the end of the K-band. The shape of the infrared energy distribution for the  metal-poor  stars  becomes  dominated  by  characteristic quasi-continuous depression attributed to pressure-induced H$_2$ absorption.

\section{Models and synthetic spectra}
\label{mod}

For this study we have used the \cite{Allard2013} BT-Settl stellar atmosphere models also used in our previous analysis of the UVES spectra of sdMs \citep{Rajpurohit2014}.  This BT-Settl model grid extends from $\teff$  = 300 to 8000 K in steps of 100 K , log\,$g$ = 2.5 to 5.5 in steps of 0.5, and [Fe/H] = -2.5 to 0.5 in steps of 0.5, accounting for alpha-enhancement. The adopted [$\alpha$/Fe] values are [M/H] $\le$ $-$1.0 then [$\alpha$/H]=+0.4, [M/H] $\ge$ $-$1.0 then [$\alpha$/H]=+0.2, and [M/H] $\ge$ $-$0.5 then [$\alpha$/H]=+0.0. These different prescriptions for $\alpha$ enhancement are rough estimates for the thin disc and thick disc \citep{Adibekyan2013}.

\begin{figure*}[!t]
\vspace*{-35mm}
	\centering
\hspace*{-2cm}	
	\includegraphics[width=19.5cm]{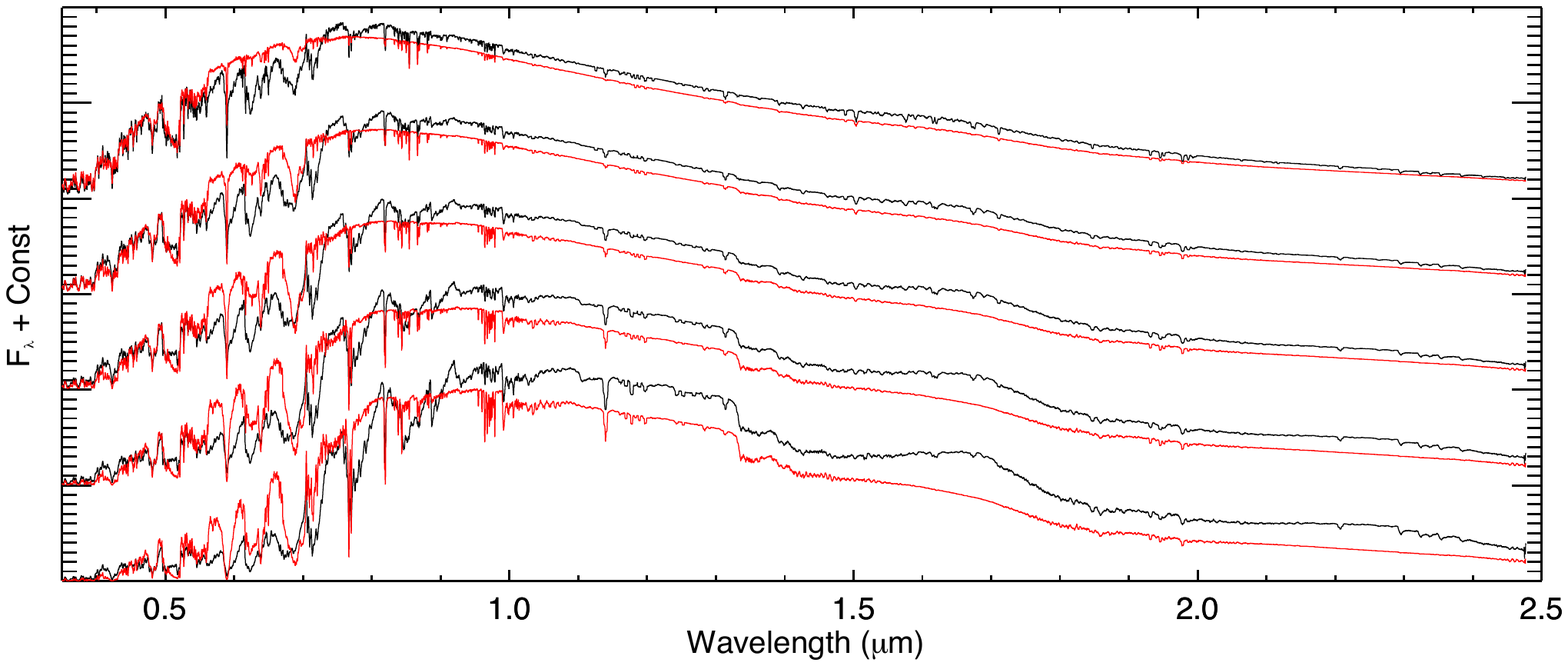}
	\hspace*{-1cm}		
        \vspace*{-35mm}
	\caption{BT-Settl synthetic spectra from 3800 K to 3000 K (top to bottom) at a step of 200 K. The black and red lines represent [Fe/H] = 0 and -2, respectivelym, for log\,$g$ = 5.0.} 
	\label{Fig3}
\end{figure*}

\begin{figure*}[!t]
\vspace*{-25mm}
	\centering
\hspace*{-2cm}	
	\includegraphics[width=19.5cm]{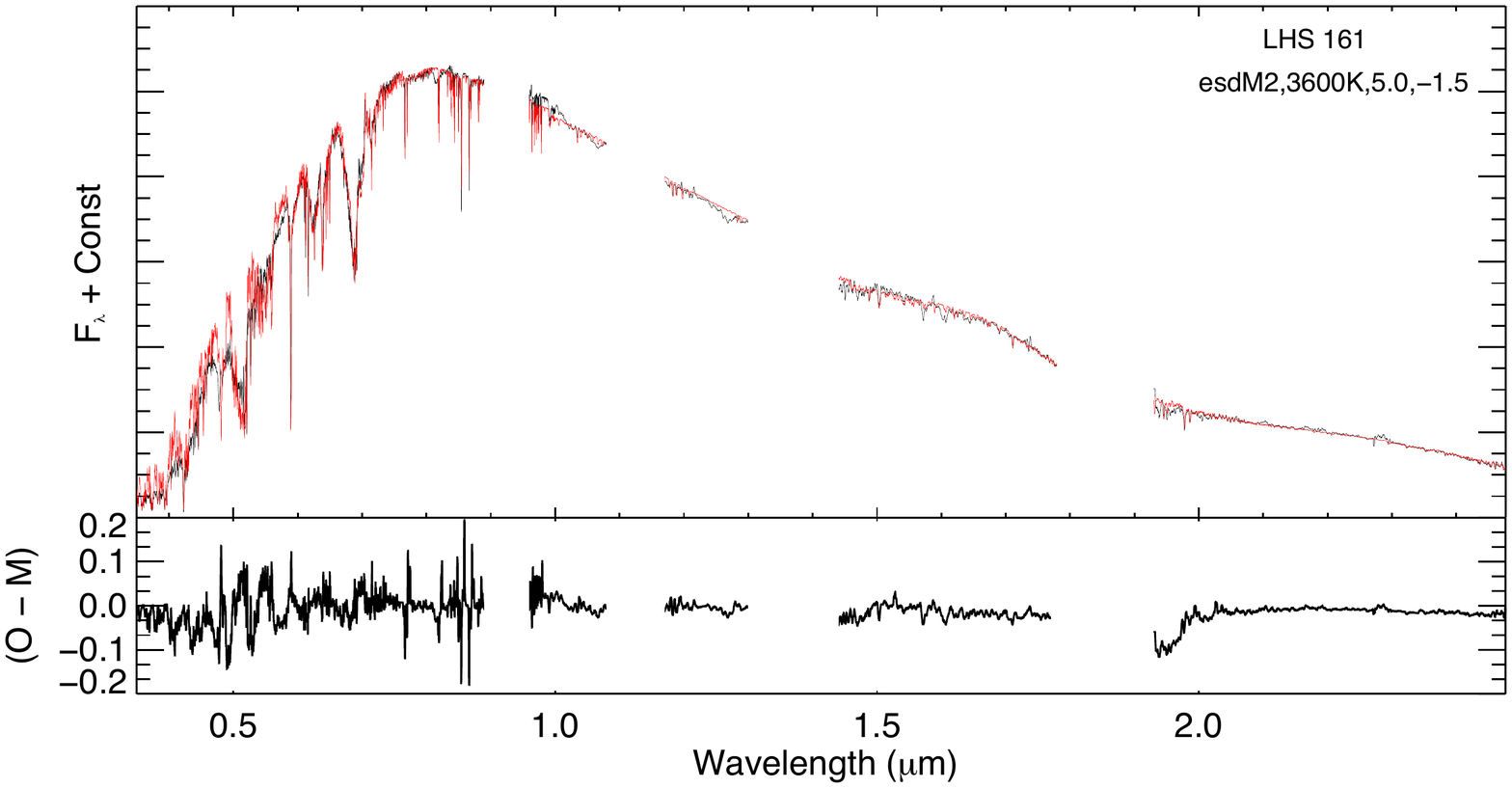}
	\hspace*{-1cm}		
        \vspace*{-30mm}
	\caption{Comparison of observed spectra of LHS 161 (esdM2) in black over the entire SED with the best-fit synthetic spectra in red (top panel). The bottom panel shows the difference in the intensities of both observed and model.} 
	\label{Fig4}
\end{figure*}

The synthetic spectra were distributed with resolution of $R \gtrsim 200\,000$ via the PHOENIX web simulator\footnote{https://phoenix.ens-lyon.fr/Grids/BT-Settl/CIFIST2011bc} and are fully described in \cite{Allard2012,Rajpurohit2012a} and \cite{Allard2013}.  Figure~\ref{Fig3} shows BT-Settl synthetic spectra with varying $\teff$  from 3800~K (top) to 3000 K (bottom) with a step of 200~K and [Fe/H] = 0.0 (black), -2.0 (red) for $\logg$= 5.0. Oxide bands, which are stronger in M-dwarf spectra, are weaker in the subdwarfs where the hydride bands dominate.

\section{Comparison with models and determination  of stellar parameters}
\label{comp}

We performed a comparison between observed and synthetic spectra. We used a least-squares minimisation program employing the new BT-Settl model atmospheres to derive fundamental stellar parameters of stars in our sample. Both theory and observation indicate that M-dwarfs have log\,$g=~5.0 \pm 0.2$ \citep{Gizis1996, Casagrande2008} except for the latest types. This is also in agreement with our previous analysis of UVES spectra for these targets, for which we derived gravities ranging from 4.5 to 5.5 using gravity sensitive atomic lines such as K~I and Na~I doublets as well as relative strength of metal hydride bands such as CaH \citep{Rajpurohit2014}.  We therefore restrict our analysis to log\,$g=~4.0~ - ~5.5$ models.

 \begin{figure*}[!thbp]
\vspace*{-15mm}
	\centering
\hspace*{-2cm}	
	\includegraphics[width=22.3cm]{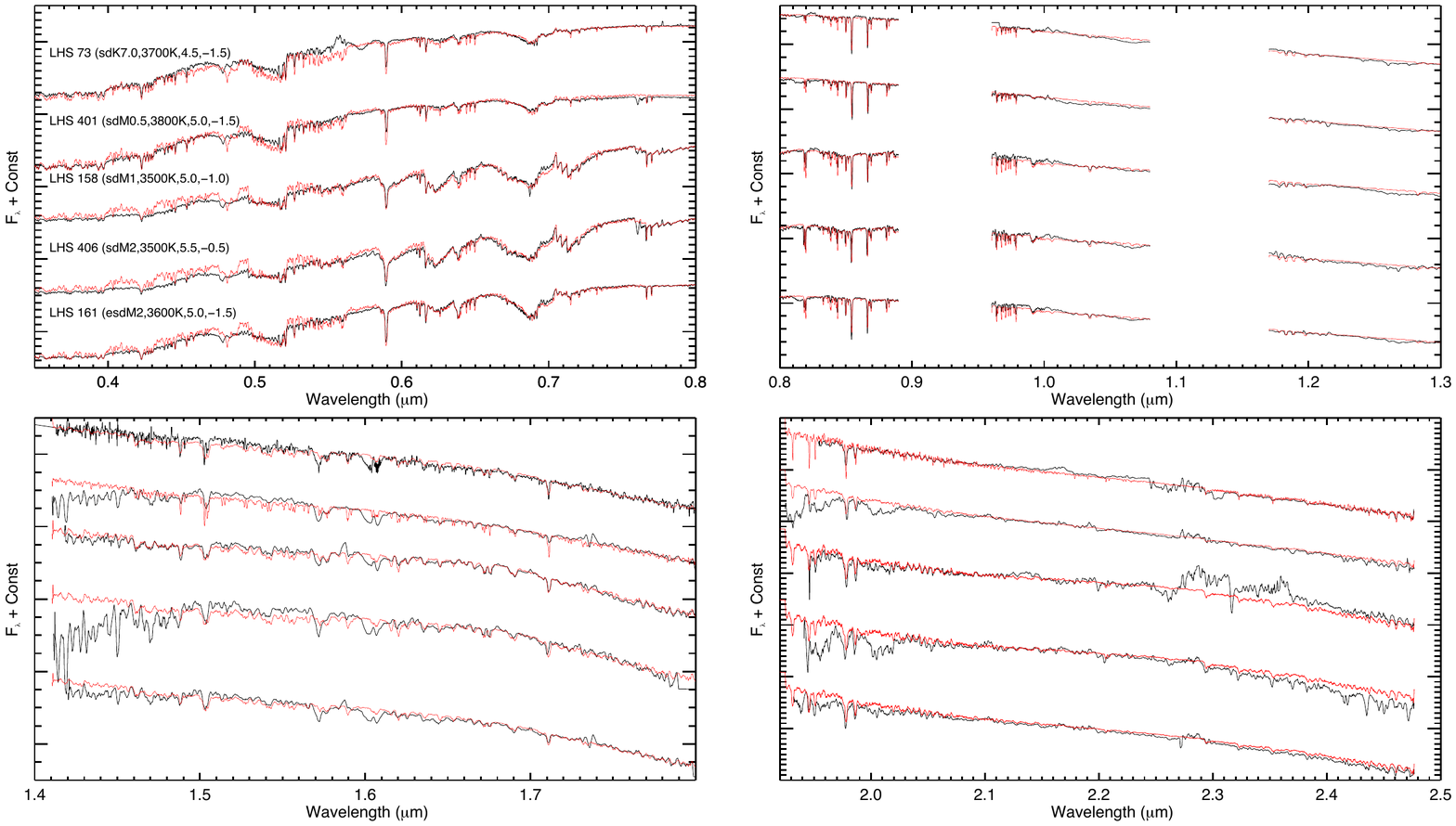}  %
	\hspace*{-1cm}		
        \vspace*{-27mm}
	\caption{X-SHOOTER spectra (black) compared with the best-fit BT-Settl synthetic spectra (red) from the spectral sequence of sdK7 to sdM2.}
	\label{Fig5}
\end{figure*}

\begin{figure*}[!thbp]
\vspace*{-15mm}
	\centering
\hspace*{-2cm}	
	\includegraphics[width=22.3cm]{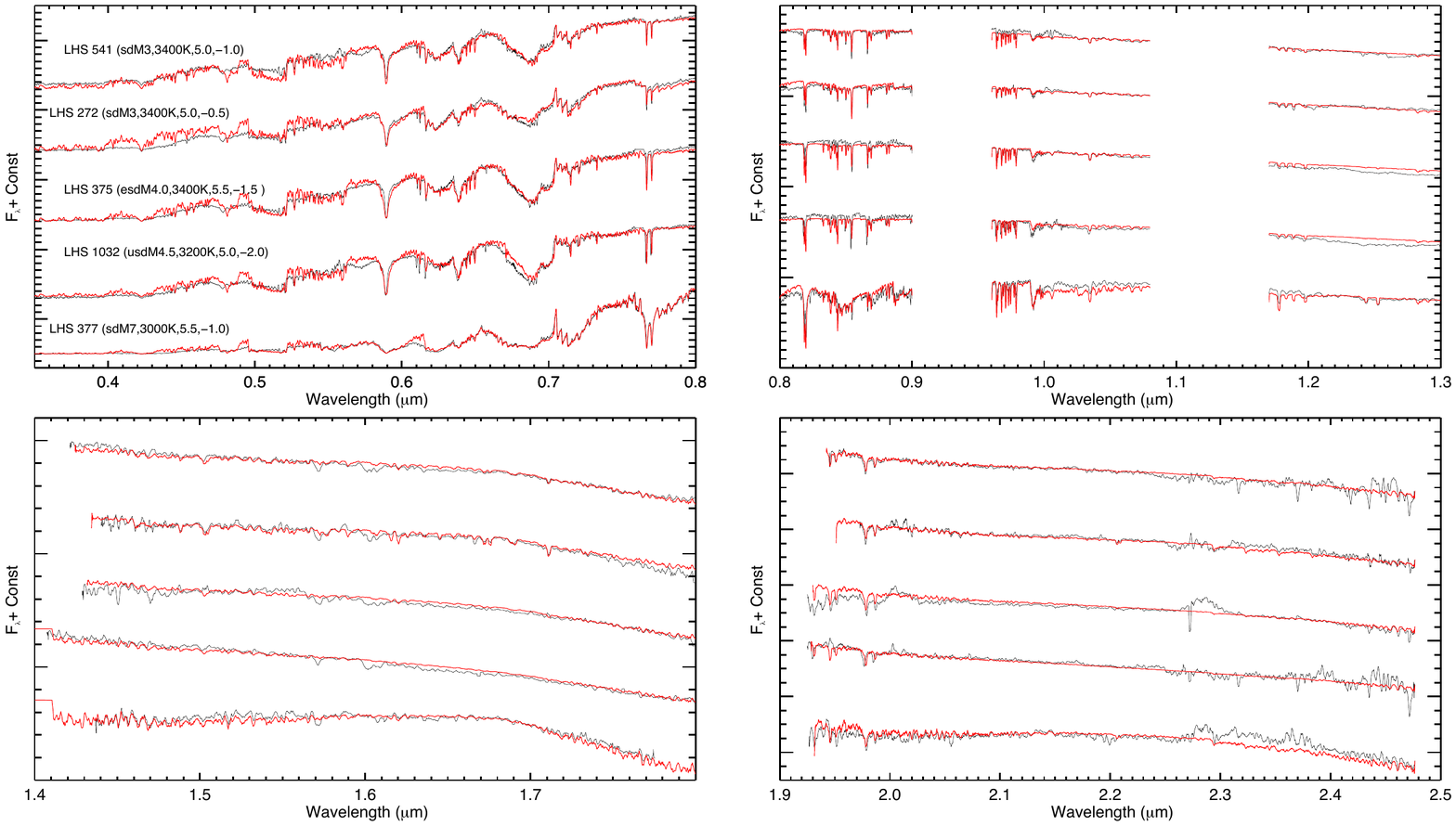}
	\hspace*{-1cm}		
\vspace*{-27mm}
	\caption{X-SHOOTER spectra (black) compared with the best-fit BT-Settl synthetic spectra (red) from the  spectral sequence of sdM3 to sdM7 which includes esdM and usdM.}
	\label{Fig6}
\end{figure*}

\begin{figure*}[!thbp]
\vspace*{-25mm}
	\centering
\hspace*{-2cm}	
\includegraphics[width=22.3cm]{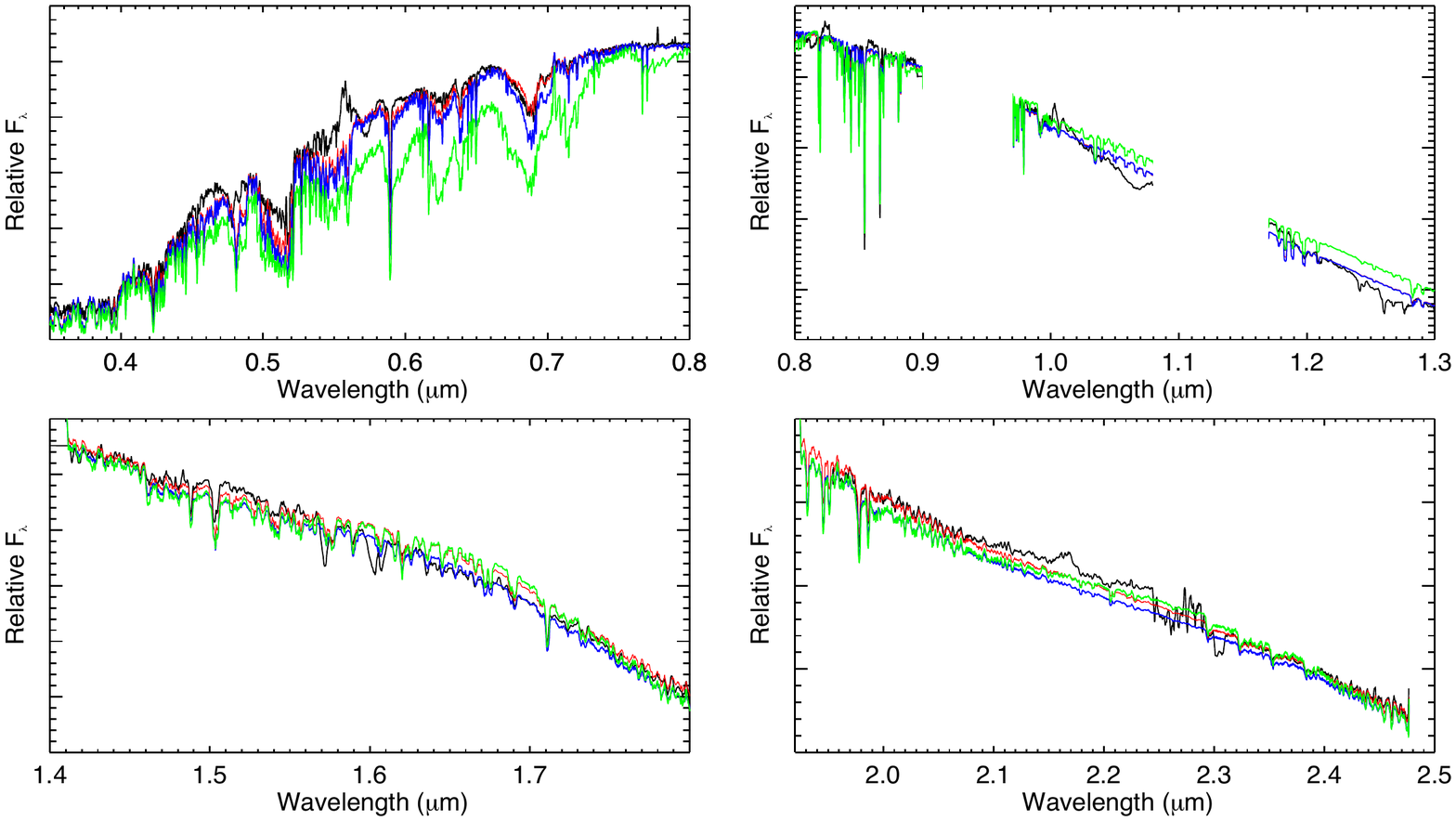}
\hspace*{-1cm}		
\vspace*{-26mm}
	\caption{X-SHOOTER spectrum LHS73 (black) compared with the best-fit BT-Settl model in red ($\teff$ = 3700 K , [Fe/H] = -1.5,  log\,$g$ = 4.5) with $\teff$ = 3700 K , [Fe/H] = -1.5,  log\,$g$ = 5.0 (blue) and $\teff$ = 3600 K , [Fe/H] = -1.5,  log\,$g$ = 5.0 (green).}
	\label{Fig7}
\end{figure*} 

\begin{figure*}[!thbp]
\vspace*{-15mm}
	\centering
\hspace*{-2cm}	
	\includegraphics[width=16.5cm,height=21cm,angle=-90]{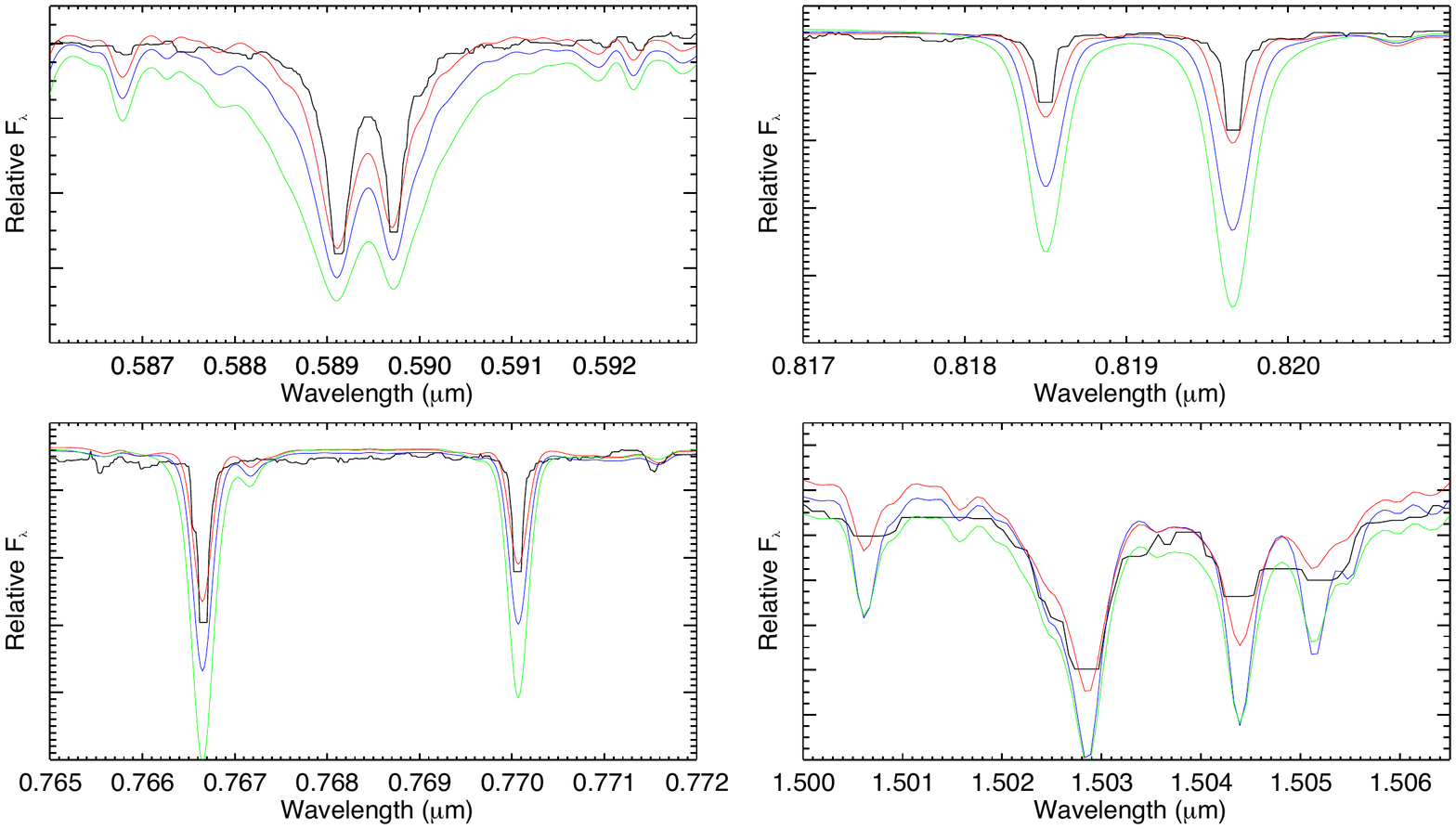}
	\hspace*{-1cm}		
\vspace*{-26mm}
	\caption{X-SHOOTER spectrum for LHS73 (black) compared with the best-fit BT-Settl model in red ($\teff$ = 3700 K , [Fe/H] = -1.5  log\,$g$ = 4.5) with log\,$g$ = 5.0 (blue) and log\,$g$ =  5.5 (green). The effect of gravity and the pressure broadening of the Na I doublet (top panel) and K I doublet (bottom panel) is clearly visible.}
	\label{Fig8}
\end{figure*}

In the process of stellar parameter determination, first, the resolution of every synthetic spectra is reduced to that of each individual observed spectrum; we normalise the spectra to unity over the entire wavelength range excluding the part with strong telluric absorption. The comparison is made using a subsample of the model atmosphere grid covering the range of 2000 K $\leq$ $\teff$  $\leq$ 4500~K, $-2.5 \leq$ [Fe/H] $\leq$ 0.5, and 4.0 $\leq$ log\,$g$ $\leq$ 5.5. Although the targets are expected to be subsolar metallicity objects, we also considered metal-rich models to demonstrate their validity. 

In the second step we interpolated the model spectra on the wavelength grid of the observed spectra and then compared each of the combined X-SHOOTER spectra to the synthetic spectra in the grid by taking the difference between the flux values of the synthetic and observed spectra at each wavelength point. The sum of the squares of the differences between model and observation is computed in the grid, and the best model for each object is selected. We selected the models that give the lowest 2-3 $\chi^2$ values. These values lie between 0.03 and 0.2. For each source and model combination the top three model spectra were decided based on the $\chi^2$ statics by considering a 3 $\sigma$ confidence level. We retain the model with the minimum $\chi^2$  values as our best-fit model. We also made a visual inspection by looking at the overall shape of the spectral energy distribution along with the line strength and shapes of various atomic species which we have used to constrained the stellar parameters such as Na I, K I, and some molecular species CaH, CaOH, CO, and TiO. We removed regions affected by telluric bands in the observed spectra in wavelength ranges from 0.9-0.95 $\mu$m, 1.1-1.17 $\mu$m,1.32-1.42 $\mu$m, and 1.78-1.95 $\mu$m. The sampling of X-SHOOTER data in the UVB (1.0”  slit width), VIS (0.9”  slit width) and NIR (0.9” slit width) arms is 6.3, 6.0, and 4.0 pix/FWHM. 

Our procedure allows a rough estimate of the probable range in the stellar parameters. The medium resolution of X-SHOOTER spectra does not provide error bars smaller than the grid spacing, which is  0.5~dex in [Fe/H] and log\,$g$  and $\pm$ 100~K in$\teff$.  We note that systematic errors due to missing or incomplete opacity sources such as TiO bands are not eliminated \citep[see latest results by][]{Baraffe2015}. However, these  uncertainties are estimated within the error bars of the values we derive for $\teff$. In figure ~\ref{Fig4} we showed the comparison of observed spectra in the middle of the $\teff$ range with the best-fit synthetic spectra over the entire SED for the direct comparison and in the bottom part of that panel a detail of the intensity difference between observed and synthetic spectra across the whole wavelength range is shown.

\section{Results}
\label{res}

We found generally good agreement of synthetic spectra with the observed spectra and conclude that our model fitting procedure can be used to estimate the effective temperature, surface gravity, and metallicity. Figure~\ref{Fig5} and Figure~\ref{Fig6} show our best model fits to the observed spectra for subdwarf and extreme subdwarf targets. Both the observed and synthetic spectra are smoothed to 20 pixels for clarity. BT-Settl models reproduce the shape of the prominent narrow atomic lines such as K~I and Na~I ; molecular bands such as CaH, CaOH, and CO ; and the pseudo-continuum formed by the TiO and water opacities from over the entire spectral region covered by the X-SHOOTER spectra. The updated molecular line list for metal hydrides and H$_2$O in the current model shows good agreement with  observations. In particular, the overall shape of the $H$ band (1.45 to 1.8 $\mu m$) and $K$  band (from 2.0 to 2.4 $\mu m$) have been improved over the previous comparison performed by \cite{Leggett2000}. The BT-Settl models also provide a fairly good match to the 0.8-1.0 $\mu m$ spectral peak, but the FeH (E $^{4}${$\Pi$}-A $^{4}${$\Pi$}) absorption \citep{Hargreaves2010} in the $H$ band is missing in the models. This could be responsible for the absorption features near 1.6 $\mu m$ seen in several of the observed spectra.

\begin{table*} [!thbp]
	\renewcommand{\arraystretch}{1.5}
	\centering
	\caption{Stellar parameters of the observed targets along with the two best-fits as determined by minimising $\chi^2$ . The uncertainty in $\teff$ is $\pm$ 100 K, whereas for $\mathrm{log}\,g$ and [Fe/H] it is $\pm$ 0.5 dex.Photometric distances were computed assuming a radius taken from 10 Gyrs isochrones from \cite{Chabrier1997} at the corresponding $\teff$}
	\begin{tabular}{ccccccccc}
		\hline
		&This study&\cite{Rajpurohit2014} &\cite{Leggett2000}&\cite{Gizis1997}&&& \\
		Target & $\teff$ /  $\mathrm{log}\,g$ / [Fe/H] &$\teff$ /  $\mathrm{log}\,g$ / [Fe/H] &$\teff$ / [Fe/H] &$\teff$ / [Fe/H] &$D_\mathrm{phot}$ (pc)&$D_\mathrm{trig}$ (pc)\\
		\hline
		
		LHS72$\dag$	&--	  			& 3900 / 4.5 / $-$1.4 &	-- &--	   &$25.6_{-10.6}^{+14.8}$	&$26.6_{-5.1}^{+8.2}$(a)\\
		LHS 73$\dag$$\dag$		&3700 / 4.5 / $-$1.5 	& 3800 / 4.5 / $-$1.4 &	-- &--	   &$9.9_{-06.2}^{+13.3}$    	&$26.6_{-5.1}^{+8.2}$(b)\\
					&3700 / 5.0 /$-$1.5	&-- &	-- &--	   &--     & --\\
		LHS 401		&3800 / 5.0 /$-$1.5	& 3800 / 4.5 / $-$1.4 &	-- &--	   &$24.4_{-08.5}^{+16.9}$    	&$26.0_{-5.2}^{+8.7}$(a)\\
					&3900 / 4.5 /$-$1.0	&-- &	-- &--	   &--     & --\\
		LHS 158		&3500 / 5.0 /$-$1.0	& 3600 / 4.5 / $-$1.0 &	-- &--	   &$22.20_{-07.6}^{+11.4}$  	& --\\
					&3600 / 4.5 /$-$1.0	&-- &	-- &--	   &--     & --\\
		LHS 406		&3500 / 5.0 /$-$1.0	& 3600 / 4.7 / $-$0.6 &	-- &--	   &$16.10_{-05.5}^{+08.3}$  	& --\\
					&3400 / 4.5 /$-$0.5	&-- &	-- &--	   &--     & --\\
		LHS 161		&3600 / 5.0 /$-$1.5	& 3700 / 4.8 / $-$1.2 &3600 / $-$1.5	  &--  &$34.30_{-09.6}^{+21.3}$  	&$39.5_{-6.5}^{+9.7}$(a)\\ 
					&3700 / 4.5 /$-$2.0	&-- &	-- &--	   &--     & --\\
		LHS 541		&3400 / 5.0 /$-$1.0	& 3500 / 5.1 / $-$1.0 &	-- 	  &3500 / $-$1.0 &$60.70_{-18.1}^{+31.2}$  	&  --\\
					&3300 / 5.0 /$-$1.0	&-- &	-- &--	   &--     & --\\
		LHS 272		&3400 / 5.0 /$-$0.5	& 3500 / 5.2 / $-$0.7 &	--   &3600 / $-$1.5 &$16.50_{-05.8}^{+09.0}$  	& -- \\
					&3500 / 5.0 /$-$0.5	&-- &	-- &--	   &--     & --\\
		LHS 375		&3400 / 5.5 /$-$1.5	& 3500 / 5.5 / $-$1.1 &3200 / $-$1.0 &3400 / $-$2.0 &$32.40_{-07.7}^{+13.3}$  	&$23.9_{-0.8}^{+0.9}$(a)\\
					&3400 / 5.0 /$-$1.5	&-- &	-- &--	   &--     & --\\
		LHS 1032		&3200 / 5.0 /$-$2.0	& 3300 / 4.5 / $-$1.7 &	-- &--	   &$61.8_{-11.0}^{+13.8}$     & --\\
					&3300 / 5.0 /$-$2.0	&-- &	-- &--	   &--     & --\\
		LHS 377		&3000 / 5.5 /$-$1.0	& 3100 / 5.3 / $-$1.0 &2900 / $-$0.5 &3200 / $-$1.5 &$37.90_{-07.7}^{+09.1}$  	&$35.2_{-0.8}^{+0.9}$(a)\\
					&3000 / 5.5 /$-$0.5	&-- &	-- &--	   &--     & --\\
		\hline
	\end{tabular}\\
	\vspace{0.25cm}
	\small{ $\dag$ The radius of LHS 72 is based on the $\teff$ given in \cite{Rajpurohit2014}}\\
	\small{ $\dag$$\dag$ LHS 72 and 73 are a common proper motion pair at the same distance}\\
	(a) \cite{Altena1995},
	(b) companion of LHS72 \citep{Reyle2006} 
\end{table*}

We have determined the parameters from the optical and from the infrared separately and obtained the same results as when using the complete X-SHOOTER SED.   We have found that the fit above 2.3~$\mu$m is not good in two cases which could be due to the low signal-to-noise ratio or to a problem with the removal of telluric or flat-fielding.   Table~2 lists the best-fit model parameters $\teff$ , log\,$g$, and [Fe/H] for each of the targets. We compare these values with those that we obtained from the analysis of the high-resolution optical  spectra obtained with UVES at VLT \citep{Rajpurohit2014}. The gravity and metallicity are in agreement, within the error bars, with the values obtained from the higher resolution UVES spectra. The effective temperature derived from the X-SHOOTER spectra are identical or systematically lower by 100~K. This could  occur because \cite{Rajpurohit2014} use only optical data in their analysis. Thus inclusion of near-infrared spectra simultaneously in the fits most likely yields lower temperatures. This has also been shown by \cite{Bayo2014}, who showed inconsistencies in the parameters derived for young M-dwarfs when using full SED, optical spectroscopy, optical + near-infrared spectra and near-infrared spectra alone.

The range of surface gravities considered in our analysis spans the entire range of expected radii from very young (a few Myrs) to the oldest M dwarfs when compared to theoretical isochrones. Unfortunately there are  no recent evolution calculations exist for the low metallicity range of our sample, but the models of \cite{Chabrier1997} predict 5.0 $\geq$ log g $\geq$ 5.5 for all M subdwarfs on the main sequence (older than $\approx 1$~Gyr). Using the isochronal radii of these evolution models, we can also calculate photometric distances (Table 2) for our sample under the assumption that they are all fully settled on the main sequence (i.e. using the 10~Gyrs isochrones). The \cite{Chabrier1997} tracks are provided for overall metallicities [Fe/H]; we have thus calculated the [Fe/H] corresponding to our [Fe/H] by accounting for the respective $\alpha$-element enhancements, and further calculated upper and lower boundaries according to the uncertainties in $\teff$ and metallicity. We assumed a total uncertainty of $^{+0.4}_{-0.5}$ in the latter since the \cite{Chabrier1997} models were calculated assuming a slightly different $\alpha$-enhancement prescription in the boundary atmospheric models of \cite{Baraffe1997}, which also used the different baseline composition of Grevesse \& Noels (1993) with higher solar CNO abundances. When comparing the solar metallicity results of \cite{Chabrier1997} to the recent \cite{Baraffe2015} models, we can also expect a moderate revision of the radii by up to $\sim$5\% if newer opacities and atmospheric boundary conditions are included.

We looked at the depth of the CO and TiO bands and matched the overall "continuum" in the optical and near-infrared regimes. For the metal-poor stars the flatness of the infrared continuum could be used to constrain the derived metallicities. Thus we find that by comparing the BT-Settl models with the spectra of M-subdwarfs over the wavelength range 0.4-2.4~$\mu$m at the resolution of the observations we are able to disentangle the effect of reduced metallicity from those of increased gravity or effective temperature which all affect the pressure structure of the photosphere. Figures~\ref{Fig7} and~\ref{Fig8} compare the sequences of these spectral models, varying $\teff$, [Fe/H] and log\,$g$. The increasing depths of the molecular bands with cooler effective temperatures are largely a consequence of the increased column depth and therefore total opacity of the associated gas species. Figure~\ref{Fig7} also compares metallicity effects. Spectral variations are far more extreme in this case. The K-band peak is suppressed at lower metallicities because of the pressure-induced H$_2$ opacities. Also, the whole structure changes as an effect of changing opacities, which can result in changing the level of the pseudo-continuum. Figure~\ref{Fig8} shows the effect of log\,$g$ at fixed $\teff$ and metallicity on the relative strength of atomic lines. The K~I doublet at 0.7665 and 0.7699 and 1.5028 $\mu$m and Na~I lines at 0.588, 0.590, 0.8185, and 0.8196 $\mu$m are particularly useful gravity discriminants for M dwarfs and subdwarfs.
			
We found that the actual fitted gravities of almost all our sample are consistent with old objects, as would be expected for such metal-poor stars, but the uncertainties of 0.5~dex do not permit a more precise estimate of their age. Their radii, however, as taken from the 10~Gyrs by \cite{Chabrier1997} isochrones, allow us to calculate photometric distances from their 2MASS magnitudes, which are listed and compared in Table~2 for all stars with trigonometric parallaxes. We found that all distances are in agreement within the relatively large uncertainties as showen in Table 2.

The gravity found for LHS~73 is surprisingly low.  The models indicate that the  CaH bands support a gravity somewhat closer to 5.0, whereas the Na~I  and K~I lines point to an even lower gravity. While no direct distance measurement exists, a parallax is available from its proper motion companion LHS~72, which has also been studied in Rajpurohit et al. (2014). It is in agreement with the photometric distance computed assuming a gravity consistent with 10~Gyr objects. When using instead the fitted gravity, i.e. assuming about twice the radius and an age < 20~Myrs, we find photometric distances of 40-50~pc for the LHS~72/73 system at their best-fit $\teff$. Thus the measured distance does lend some support to higher surface gravities. Furthermore, the proper motion of the pair indicates a tangential velocity of at least 275 km/s (for the lower limit of the distance) and thus a clear old population dynamics, making such a young age extremely unlikely. Thus the discrepancy to the fitted gravity remains an issue and is probably an indication that some details of the line formation are not modelled well, or is due to the neglect of NLTE effects in the BT-Settl model grid used. We have also found differences with the NextGen models used by \cite{Leggett2000} in the stellar parameters when comparing them with earlier studies on common objects (Table~2). This shows the importance of the model used -- the NextGen model includes incomplete TiO and water vapour line list, which could lead to uncertainty in both $\teff$ and [Fe/H] determinations -- and of the wavelength range studied.

\section{Conclusion}
\label{ccl}

In this paper, we obtained 0.3-2.5 $\mu$m simultaneous X-SHOOTER spectra for ten subdwarfs including extreme and ultra-subdwarfs, and described the results of their comparison to the BT-Settl synthetic spectra. We found that the parameters of an observed star are best determined by using a combination of both low-resolution and high-resolution spectra in order to disentangle the atmospheric parameters $\teff$, log\,$g$, and [Fe/H] which can have compensating effects.  

Models at the low metallicity used in this analysis have not been extensively tested before over full wavelength coverage for M-subdwarfs.  This model grid constitutes a significant improvement over the earlier models by \cite{Allard1990,Allard1998} and \cite{Allard2000} owing to more complete and accurate line lists for  TiO and  H$_2$O, to the current computing possibility to include all the lines in the radiative transfer, and to the revised solar abundances. The synthetic spectra demonstrate the influence of the metallicity and gravity. We found that gravity has a relatively small influence on the spectra and the overall energy distribution, as was also found with earlier models \citep{Leggett1998,Leggett2000}, but it has a significant effect on high-resolution line profiles and details of the band systems. However, the effects of gravity become stronger with lower effective temperatures. The metallicity has, on the other hand, a large effect on the spectra. We determined stellar parameters and found a good agreement with those derived from higher spectral resolution observations in the optical wavelength \citep{Rajpurohit2014}. The difference between the parameters determined from UVES spectra and from X-SHOOTER spectra agree within the error bars defined by the grid spacing. The disentangling between the effects of temperature, metallicity and gravity is possible because the models show a large variation in their spectral features due to either molecular or atomic opacities, and the parameters are obtained with a high confidence level despite the existing grid spacing. Features formed by the CIA bands are reproduced well enough  through the NIR continuum. 

The overall agreement of models with the observations is much improved when compared to the earlier work. \cite{Gizis1997} obtained values for $\teff$ and [Fe/H] of sdMs and esdMs by comparing optical spectra in the region from 6200-7400 $\AA$ (resolution = 3-4 $\AA$). \cite{Gizis1997} used the extended model grid \citep{Allard1995} which assumed local thermodynamic equilibrium (LTE) for their comparison with observed spectra. \cite{Leggett1996,Leggett1998,Leggett2000} compared spectra in the range from 0.6 - 2.5  $\mu$m with the NextGen \citep{Hauschildt1999} model grid. The BT-Settl models used in this paper also assume LTE and resolve the historical discrepancy in the stellar parameters obtained from the infrared and optical spectral regions also seen by \cite{Viti1997}.  

Even though the BT-Settl collection does a better job in reproducing simultaneously the optical and near-infrared features of these cool metal-poor sources, there is still room for improvement since there are regions where the fit is not optimal, in particular below 0.45 $\mu$m and in the H and K bands. This can be due to missing CaOH bands in the V bandpass in the models. Also, a complete FeH line list is currently missing in the H bandpass. An accurate and complete TiO line list is currently being developed by the ExoMol group. Upgrading these opacities is the next step in improving these models in the near future, before computing detailed model atmosphere grids and interior and evolution models at finer steps in the atmospheric parameters.  We also note that at low metallicity, the effects of temperature inhomogeneities in the atmosphere begin to have greater impact on the spectrum formation, which can only be accurately modelled with 3D RHD simulations and ultimately 3D radiative transfer. 

\begin{acknowledgements}
We thank the referee S. Leggett for her useful comments and suggestion on the paper. We acknowledge observing support from the ESO staff. The research leading to these results has received funding from the French Programme National de Physique Stellaire and the Programme National de Planetologie of CNRS (INSU). The computations were performed at the {\sl P\^ole Scientifique de Mod\'elisation Num\'erique} (PSMN) at the {\sl \'Ecole Normale Sup\'erieure} (ENS) in Lyon, and at the {\sl Gesellschaft f{\"u}r Wissenschaftliche Datenverarbeitung G{\"o}ttingen} in collaboration with the Institut f{\"u}r Astrophysik G{\"o}ttingen. D. H.  acknowledges support from the European Research Council under the European Community’s Seventh Framework Programme (FP7/2007-2013 Grant Agreement No. 247060) and from the Collaborative Research Centre SFB 881 ``The Milky Way System'' (subproject A4) of the German Research Foundation (DFG).  O.M. acknowledges support from CNES. This work has been partly carried out thanks to the support of the A*MIDEX project (n\textsuperscript{o} ANR-11-IDEX-0001-02) funded by the ``Investissements d'Avenir'' French Government programme, managed by the French National Research Agency (ANR). J.G.F-T is currently supported by Centre National d'Etudes Spatiales (CNES) through PhD grant 0101973 and the R\'egion de Franche-Comt\'e and by the French Programme National de Cosmologie et Galaxies (PNCG). A. B. acknowledges financial support from the Proyecto Fondecyt de Iniciaci\'on 11140572.

\end{acknowledgements}

\bibliographystyle{aa}
\bibliography{ref}

\begin{thebibliography}{50}
\expandafter\ifx\csname natexlab\endcsname\relax\def\natexlab#1{#1}\fi

\bibitem[{{Adibekyan} {et~al.}(2013){Adibekyan}, {Figueira}, {Santos},
  {Hakobyan}, {Sousa}, {Pace}, {Delgado Mena}, {Robin}, {Israelian}, \&
  {Gonz{\'a}lez Hern{\'a}ndez}}]{Adibekyan2013}
{Adibekyan}, V.~Z., {Figueira}, P., {Santos}, N.~C., {et~al.} 2013, \aap, 554,
  A44

\bibitem[{{Allard}(1990)}]{Allard1990}
{Allard}, F. 1990, PhD thesis, PhD thesis.~Ruprecht Karls Univ.~Heidelberg,
  (1990)

\bibitem[{{Allard} {et~al.}(1998){Allard}, {Alexander}, \&
  {Hauschildt}}]{Allard1998}
{Allard}, F., {Alexander}, D.~R., \& {Hauschildt}, P.~H. 1998, in Astronomical
  Society of the Pacific Conference Series, Vol. 154, Cool Stars, Stellar
  Systems, and the Sun, ed. {R.~A.~Donahue \& J.~A.~Bookbinder}, 63--+

\bibitem[{{Allard} \& {Hauschildt}(1995)}]{Allard1995}
{Allard}, F. \& {Hauschildt}, P.~H. 1995, \apj, 445, 433

\bibitem[{{Allard} {et~al.}(1997){Allard}, {Hauschildt}, {Alexander}, \&
  {Starrfield}}]{Allard1997}
{Allard}, F., {Hauschildt}, P.~H., {Alexander}, D.~R., \& {Starrfield}, S.
  1997, \araa, 35, 137

\bibitem[{{Allard} {et~al.}(2001){Allard}, {Hauschildt}, {Alexander},
  {Tamanai}, \& {Schweitzer}}]{Allard2001}
{Allard}, F., {Hauschildt}, P.~H., {Alexander}, D.~R., {Tamanai}, A., \&
  {Schweitzer}, A. 2001, \apj, 556, 357

\bibitem[{{Allard} {et~al.}(2000){Allard}, {Hauschildt}, \&
  {Schwenke}}]{Allard2000}
{Allard}, F., {Hauschildt}, P.~H., \& {Schwenke}, D. 2000, \apj, 540, 1005

\bibitem[{{Allard} {et~al.}(2012){Allard}, {Homeier}, \&
  {Freytag}}]{Allard2012}
{Allard}, F., {Homeier}, D., \& {Freytag}, B. 2012, Royal Society of London
  Philosophical Transactions Series A, 370, 2765

\bibitem[{{Allard} {et~al.}(2013){Allard}, {Homeier}, {Freytag},
  {Schaffenberger}, {}, \& {Rajpurohit}}]{Allard2013}
{Allard}, F., {Homeier}, D., {Freytag}, B., {et~al.} 2013, Memorie della
  Societa Astronomica Italiana Supplementi, 24, 128

\bibitem[{{Asplund} {et~al.}(2009){Asplund}, {Grevesse}, {Sauval}, \&
  {Scott}}]{Asplund2009}
{Asplund}, M., {Grevesse}, N., {Sauval}, A.~J., \& {Scott}, P. 2009, \araa, 47,
  481

\bibitem[{{Baraffe} {et~al.}(1997){Baraffe}, {Chabrier}, {Allard}, \&
  {Hauschildt}}]{Baraffe1997}
{Baraffe}, I., {Chabrier}, G., {Allard}, F., \& {Hauschildt}, P.~H. 1997, \aap,
  327, 1054

\bibitem[{{Baraffe} {et~al.}(2015){Baraffe}, {Homeier}, {Allard}, \&
  {Chabrier}}]{Baraffe2015}
{Baraffe}, I., {Homeier}, D., {Allard}, F., \& {Chabrier}, G. 2015, ArXiv
  e-prints

\bibitem[{{Bayo} {et~al.}(2012){Bayo}, {Barrado}, {Hu{\'e}lamo},
  {Morales-Calder{\'o}n}, {Melo}, {Stauffer}, \& {Stelzer}}]{Bayo2012}
{Bayo}, A., {Barrado}, D., {Hu{\'e}lamo}, N., {et~al.} 2012, \aap, 547, A80

\bibitem[{{Bayo} {et~al.}(2011){Bayo}, {Barrado}, {Stauffer},
  {Morales-Calder{\'o}n}, {Melo}, {Hu{\'e}lamo}, {Bouy}, {Stelzer}, {Tamura},
  \& {Jayawardhana}}]{Bayo2011}
{Bayo}, A., {Barrado}, D., {Stauffer}, J., {et~al.} 2011, \aap, 536, A63

\bibitem[{{Bayo} {et~al.}(2014){Bayo}, {Rodrigo}, {Barrado}, \&
  {Allard}}]{Bayo2014}
{Bayo}, A., {Rodrigo}, C., {Barrado}, D., \& {Allard}, F. 2014, \memsai, 85,
  773

\bibitem[{{Bochanski} {et~al.}(2010){Bochanski}, {Hawley}, {Covey}, {West},
  {Reid}, {Golimowski}, \& {Ivezi{\'c}}}]{Bochanski2010}
{Bochanski}, J.~J., {Hawley}, S.~L., {Covey}, K.~R., {et~al.} 2010, \aj, 139,
  2679

\bibitem[{{Brott} \& {Hauschildt}(2005)}]{Brott2005}
{Brott}, I. \& {Hauschildt}, P.~H. 2005, in ESA Special Publication, Vol. 576,
  The Three-Dimensional Universe with Gaia, ed. C.~{Turon}, K.~S. {O'Flaherty},
  \& M.~A.~C. {Perryman}, 565

\bibitem[{{Burgasser}(2002)}]{Burgasser2002}
{Burgasser}, A.~J. 2002, PhD thesis, CALIFORNIA INSTITUTE OF TECHNOLOGY

\bibitem[{{Caffau} {et~al.}(2011){Caffau}, {Ludwig}, {Steffen}, {Freytag}, \&
  {Bonifacio}}]{Caffau2011}
{Caffau}, E., {Ludwig}, H.-G., {Steffen}, M., {Freytag}, B., \& {Bonifacio}, P.
  2011, \solphys, 268, 255

\bibitem[{{Casagrande} {et~al.}(2008){Casagrande}, {Flynn}, \&
  {Bessell}}]{Casagrande2008}
{Casagrande}, L., {Flynn}, C., \& {Bessell}, M. 2008, \mnras, 389, 585

\bibitem[{{Chabrier} \& {Baraffe}(1997)}]{Chabrier1997}
{Chabrier}, G. \& {Baraffe}, I. 1997, \aap, 327, 1039

\bibitem[{{Dawson} \& {De Robertis}(2000)}]{Dawson2000}
{Dawson}, P.~C. \& {De Robertis}, M.~M. 2000, \aj, 120, 1532

\bibitem[{{Freudling} {et~al.}(2013){Freudling}, {Romaniello}, {Bramich},
  {Ballester}, {Forchi}, {Garc{\'{\i}}a-Dabl{\'o}}, {Moehler}, \&
  {Neeser}}]{Freudling2013}
{Freudling}, W., {Romaniello}, M., {Bramich}, D.~M., {et~al.} 2013, \aap, 559,
  A96

\bibitem[{{Geballe} {et~al.}(1996){Geballe}, {Kulkarni}, {Woodward}, \&
  {Sloan}}]{Geballe1996}
{Geballe}, T.~R., {Kulkarni}, S.~R., {Woodward}, C.~E., \& {Sloan}, G.~C. 1996,
  \apjl, 467, L101

\bibitem[{{Gizis}(1996)}]{Gizis1996}
{Gizis}, J.~E. 1996, in Astronomical Society of the Pacific Conference Series,
  Vol. 109, Cool Stars, Stellar Systems, and the Sun, ed. {R.~Pallavicini \&
  A.~K.~Dupree}, 683

\bibitem[{{Gizis}(1997)}]{Gizis1997}
{Gizis}, J.~E. 1997, \aj, 113, 806

\bibitem[{{Hargreaves} {et~al.}(2010){Hargreaves}, {Hinkle}, {Bauschlicher},
  {Wende}, {Seifahrt}, \& {Bernath}}]{Hargreaves2010}
{Hargreaves}, R.~J., {Hinkle}, K.~H., {Bauschlicher}, Jr., C.~W., {et~al.}
  2010, \aj, 140, 919

\bibitem[{{Hauschildt} {et~al.}(1999){Hauschildt}, {Allard}, \&
  {Baron}}]{Hauschildt1999}
{Hauschildt}, P.~H., {Allard}, F., \& {Baron}, E. 1999, \apj, 512, 377

\bibitem[{{Helling} {et~al.}(2008){Helling}, {Ackerman}, {Allard}, {Dehn},
  {Hauschildt}, {Homeier}, {Lodders}, {Marley}, {Rietmeijer}, {Tsuji}, \&
  {Woitke}}]{Helling2008a}
{Helling}, C., {Ackerman}, A., {Allard}, F., {et~al.} 2008, \mnras, 391, 1854

\bibitem[{{Jao} {et~al.}(2008){Jao}, {Henry}, {Beaulieu}, \&
  {Subasavage}}]{Jao2008}
{Jao}, W.-C., {Henry}, T.~J., {Beaulieu}, T.~D., \& {Subasavage}, J.~P. 2008,
  \aj, 136, 840

\bibitem[{{Jones} {et~al.}(1994){Jones}, {Longmore}, {Jameson}, \&
  {Mountain}}]{Jones1994}
{Jones}, H.~R.~A., {Longmore}, A.~J., {Jameson}, R.~F., \& {Mountain}, C.~M.
  1994, \mnras, 267, 413

\bibitem[{{Kirkpatrick} {et~al.}(1993){Kirkpatrick}, {Kelly}, {Rieke},
  {Liebert}, {Allard}, \& {Wehrse}}]{Kirkpatrick1993}
{Kirkpatrick}, J.~D., {Kelly}, D.~M., {Rieke}, G.~H., {et~al.} 1993, \apj, 402,
  643

\bibitem[{{Leggett} {et~al.}(1996){Leggett}, {Allard}, {Berriman}, {Dahn}, \&
  {Hauschildt}}]{Leggett1996}
{Leggett}, S.~K., {Allard}, F., {Berriman}, G., {Dahn}, C.~C., \& {Hauschildt},
  P.~H. 1996, \apjs, 104, 117

\bibitem[{{Leggett} {et~al.}(2000){Leggett}, {Allard}, {Dahn}, {Hauschildt},
  {Kerr}, \& {Rayner}}]{Leggett2000}
{Leggett}, S.~K., {Allard}, F., {Dahn}, C., {et~al.} 2000, \apj, 535, 965

\bibitem[{{Leggett} {et~al.}(2001){Leggett}, {Allard}, {Geballe}, {Hauschildt},
  \& {Schweitzer}}]{Leggett2001}
{Leggett}, S.~K., {Allard}, F., {Geballe}, T.~R., {Hauschildt}, P.~H., \&
  {Schweitzer}, A. 2001, \apj, 548, 908

\bibitem[{{Leggett} {et~al.}(1998){Leggett}, {Allard}, \&
  {Hauschildt}}]{Leggett1998}
{Leggett}, S.~K., {Allard}, F., \& {Hauschildt}, P.~H. 1998, \apj, 509, 836

\bibitem[{{L{\'e}pine} {et~al.}(2007){L{\'e}pine}, {Rich}, \&
  {Shara}}]{Lepine2007}
{L{\'e}pine}, S., {Rich}, R.~M., \& {Shara}, M.~M. 2007, \apj, 669, 1235

\bibitem[{{Neves} {et~al.}(2014){Neves}, {Bonfils}, {Santos}, {Delfosse},
  {Forveille}, {Allard}, \& {Udry}}]{Neves2014}
{Neves}, V., {Bonfils}, X., {Santos}, N.~C., {et~al.} 2014, \aap, 568, A121

\bibitem[{{Rajpurohit} {et~al.}(2013){Rajpurohit}, {Reyl{\'e}}, {Allard},
  {Homeier}, {Schultheis}, {Bessell}, \& {Robin}}]{Rajpurohit2013}
{Rajpurohit}, A.~S., {Reyl{\'e}}, C., {Allard}, F., {et~al.} 2013, \aap, 556,
  A15

\bibitem[{{Rajpurohit} {et~al.}(2014){Rajpurohit}, {Reyl{\'e}}, {Allard},
  {Scholz}, {Homeier}, {Schultheis}, \& {Bayo}}]{Rajpurohit2014}
{Rajpurohit}, A.~S., {Reyl{\'e}}, C., {Allard}, F., {et~al.} 2014, \aap, 564,
  A90

\bibitem[{{Rajpurohit} {et~al.}(2012){Rajpurohit}, {Reyl{\'e}}, {Schultheis},
  {Leinert}, {Allard}, {Homeier}, {Ratzka}, {Abraham}, {Moster}, {Witte}, \&
  {Ryde}}]{Rajpurohit2012a}
{Rajpurohit}, A.~S., {Reyl{\'e}}, C., {Schultheis}, M., {et~al.} 2012, \aap,
  545, A85

\bibitem[{{Reid} \& {Hawley}(2005)}]{Reid2005b}
{Reid}, I.~N. \& {Hawley}, S.~L. 2005, {New light on dark stars : red dwarfs,
  low-mass stars, brown dwarfs}

\bibitem[{{Reyl{\'e}} {et~al.}(2006){Reyl{\'e}}, {Scholz}, {Schultheis},
  {Robin}, \& {Irwin}}]{Reyle2006}
{Reyl{\'e}}, C., {Scholz}, R.-D., {Schultheis}, M., {Robin}, A.~C., \& {Irwin},
  M. 2006, \mnras, 373, 705

\bibitem[{{Ryan} \& {Norris}(1991)}]{Ryan1991b}
{Ryan}, S.~G. \& {Norris}, J.~E. 1991, \aj, 101, 1835

\bibitem[{{Ryan} {et~al.}(1991){Ryan}, {Norris}, \& {Bessell}}]{Ryan1991a}
{Ryan}, S.~G., {Norris}, J.~E., \& {Bessell}, M.~S. 1991, \aj, 102, 303

\bibitem[{{Savcheva} {et~al.}(2014){Savcheva}, {West}, \&
  {Bochanski}}]{Savcheva2014}
{Savcheva}, A.~S., {West}, A.~A., \& {Bochanski}, J.~J. 2014, \apj, 794, 145

\bibitem[{{Tsuji} {et~al.}(1996{\natexlab{a}}){Tsuji}, {Ohnaka}, \&
  {Aoki}}]{Tsuji1996b}
{Tsuji}, T., {Ohnaka}, K., \& {Aoki}, W. 1996{\natexlab{a}}, \aap, 305, L1+

\bibitem[{{Tsuji} {et~al.}(1996{\natexlab{b}}){Tsuji}, {Ohnaka}, {Aoki}, \&
  {Nakajima}}]{Tsuji1996a}
{Tsuji}, T., {Ohnaka}, K., {Aoki}, W., \& {Nakajima}, T. 1996{\natexlab{b}},
  \aap, 308, L29

\bibitem[{{van Altena} {et~al.}(1995){van Altena}, {Lee}, \&
  {Hoffleit}}]{Altena1995}
{van Altena}, W.~F., {Lee}, J.~T., \& {Hoffleit}, D. 1995, VizieR Online Data
  Catalog, 1174, 0

\bibitem[{{Viti} {et~al.}(1997){Viti}, {Jones}, {Schweitzer}, {Allard},
  {Hauschildt}, {Tennyson}, {Miller}, \& {Longmore}}]{Viti1997}
{Viti}, S., {Jones}, H.~R.~A., {Schweitzer}, A., {et~al.} 1997, \mnras, 291,
  780

\end{thebibliography}
\end{document}